\documentclass[preprint]{revtex4}
\usepackage{color}

\begin{document}

\baselineskip = 1.1 \baselineskip
\oddsidemargin=10mm
\parindent=1cm

\vspace{1.0cm}

\title
{\Large {Critical exponents and the pseudo-$\epsilon$ expansion}}

\author{M. A. Nikitina$^{1,2}$, A. I. Sokolov$^1$}

\email{ais2002@mail.ru}
\address
{$^1$ St. Petersburg State University, St. Petersburg, Russia}
\address
{$^2$ St. Petersburg National Research University ITMO, St. Petersburg,
Russia}

\date{\today}

\begin{abstract}
We present the pseudo-$\epsilon$ expansions ($\tau$-series) for the
critical exponents of a $\lambda\phi^4$ three-dimensional $O(n)$-symmetric
model obtained on the basis of six-loop renormalization-group expansions.
Concrete numerical results are presented for physically interesting cases
$n = 1$, $n = 2$, $n = 3$ and $n = 0$, as well as for $4 \le n \le 32$ in
order to clarify the general properties of the obtained series. The
pseudo-$\epsilon$-expansions for the exponents $\gamma$ è $\alpha$ have
small and rapidly decreasing coefficients. So, even the direct summation
of the $\tau$-series leads to fair estimates for critical exponents, while
addressing Pade approximants enables one to get high-precision numerical
results. In contrast, the coefficients of the pseudo-$\epsilon$ expansion
of the scaling correction exponent $\omega$ do not exhibit any tendency to
decrease at physical values of $n$. But the corresponding series are
sign-alternating, and to obtain reliable numerical estimates, it also
suffices to use simple Pad\'e approximants in this case. The
pseudo-$\epsilon$ expansion technique can therefore be regarded as a
specific resummation method converting divergent renormalization-group
series into expansions that are computationally convenient.

\end{abstract}


\maketitle

\section{Introduction}

\vspace{0.5cm}

\textwidth=15.5cm

Although no exact solution of the problem of the phase transition in
the three-dimensional $n$-vector model has yet been obtained, the
critical exponents of systems described by this model (Heisenberg and
uniaxial ferromagnets, Bose superfluids, etc.) have been calculated
with a rather high accuracy. One of the most efficient methods, which
allows obtaining precise quantitative results, is the field-theory renormalization
group (RG) method. The calculations of five-, six-, and seven-loop RG expansions
\cite{BNGM76,Guelph,BNM78,MN91,AS95} showed the way for calculating the
critical exponents, the ratios of critical amplitudes, and other universal
characteristics of the critical behavior of three-dimensional systems
with an absolute error less than or equal to 0.002 to 0.003
\cite{BNM78,LGZ80,GZJ98,Kl99,Susl08}.

The field-theory RG technique is based on the renormalized perturbation
theory, i.e., on a regular mathematical procedure that allows obtaining
the critical exponents and other observables as power series in the effective
dimensionless coupling constant (renormalized charge) $g$. The asymptotic
value of $g$ is a nontrivial zero of the Gell-Mann–-Low function $\beta(g)$,
which can also be calculated by the perturbation theory methods. But the
perturbation series, as is known, diverge, and the expansion parameter is not
small for models with spaces of physical dimension ($D = 3, D = 2$)), i.e.,
$g\sim1\div2$. To obtain quantitative results, we must therefore use different
methods for summing the divergent series (see, e.g., \cite{ZJ01,ZJ,KS01,PV02,Susl05}).
When using an alternative approach based on $(4-\epsilon)$-dimensional models, we
must use resummation procedures because the $\epsilon$ expansions have the same
drawbacks in the physical limit $\epsilon \to 1$ $\epsilon$ as the expansions
in the charge.

At the same time, there is a method for transforming the initial
RG expansions into series with small coefficients rapidly decreasing
in absolute value. We mean the pseudo-$\epsilon$ expansion method
proposed by Nickel (see reference [19] in \cite{LGZ80}). The main idea
of the method is to replace the coefficient of the linear term in the
expansion of the $\beta$ function with a fictitious small parameter $\tau$,
to calculate the Wilson fixed-point coordinate as a power series in $\tau$,
and to construct the $\tau$-expansions of the critical exponents and other
universal quantities. The structure of the series obtained with this
technique turns out to be very convenient computationally, and reliable
numerical results can hence be obtained by directly summing these series or
by using simple Pad\'e approximants \cite{FH97,FH99,FHY2000,HID04,CP05,NS14c,NS14e}.
The pseudo-$\epsilon$ expansion method proved to be highly efficient even in
the case of two-dimensional systems, where the known RG expansions are shorter
and diverge faster than in the case of three-dimensional models \cite{COPS04,S05,S13,NS13}.

Our goal here is to present pseudo-$\epsilon$ expansions of critical exponents
of a three-dimensional $n$-vector model, to analyze their structure for
different values of $n$, and to discuss the numerical results obtained in
the framework of this approach. We mainly focus on the cases $n = 1$, $n = 2$,
$n = 3$, and $n = 0$, which are physically the most interesting. We obtain
numerical estimates by using Pad\'e approximants and by summing the obtained
pseudo-$\epsilon$ expansions directly.

\section{Pseudo-$\epsilon$ expansions of critical exponents for arbitrary $n$}

The critical thermodynamics of three-dimensional systems with an $O(n)$-symmetric
order parameter are described by the Euclidean field theory with a
$\lambda\phi^4$ type interaction whose Hamiltonian has the form

\begin{equation}
\label{eq:1} H = \int d^{3}x \Biggl[{1 \over 2}( m_0^2 \varphi_{\alpha}^2
 + (\nabla \varphi_{\alpha})^2)
+ {\lambda \over 24} (\varphi_{\alpha}^2)^2 \Biggr] ,
\end{equation}
where the squared bare mass $m_0$ is proportional to $T - T_c^{(0)}$ and
$T_c^{(0)}$ is the temperature of the phase transition without fluctuations.
The RG expansions of the $\beta$ function and the critical exponents, obtained
in the framework of the massive theory with a propagator, a vertex function,
and a "three-leg" function $\Gamma_R^{1,2}$ normalized at zero external momenta,
\begin{eqnarray}
\label{eq:2} G_R^{-1} (0, m, g_4) = m^2, \qquad \quad {{\partial G_R^{-1}
(p, m, g_4)} \over {\partial p^2}}
\bigg\arrowvert_{p^2 = 0} = 1 , \\
\nonumber \Gamma_R (0, 0, 0, m, g) = m^2 g_4, \qquad \quad \Gamma_R^{1,2}
(0, 0, m, g_4) = 1,
\end{eqnarray}
are currently known in the six-loop approximation \cite{Guelph,BNM78,AS95}.
To obtain the sought pseudo-$\epsilon$ expansions ($\tau$-series), it
suffices to substitute the $\tau$-series for the Wilson fixed point
coordinate \cite{NS13} in the RG expansions of the critical exponents
and to reexpand them in power series in $\tau$. Here, we present
calculation results for the critical exponents of the susceptibility
and heat capacity because these exponents are measured in experiments
most frequently and most precisely. The pseudo-$\epsilon$ expansions
of other critical exponents can be obtained based on the $\tau$-series
for the parameters $\gamma$ and $\alpha$ using the well-known scaling
relations. Hence, we have
\begin{eqnarray}
\label{gamma-tau} \gamma &=& 1 + {(n + 2) \tau \over 2(n + 8)} + {\tau^2
(n + 2)\over 108(n + 8)^3} \Biggl(27 n^2 + 490 n + 1088 \Biggr)
\nonumber \\
&+& {\tau^3 (n + 2)\over 8(n + 8)^5} \biggl(n^4 + 37.936996 n^3 +
428.99211 n^2 + 1073.0522 n - 199.53302 \biggr)
\nonumber \\
&+& {\tau^4 (n + 2) \over 16(n + 8)^7} \biggl(n^6 + 58.533991 n^5 +
1336.0645 n^4 + 13115.226 n^3
\nonumber \\
&+& 46827.023 n^2 + 60693.508 n + 45180.873 \biggr)
\nonumber \\
&+& {\tau^5 (n + 2) \over 32(n + 8)^9} \biggl(n^8 + 79.88230 n^7 +
2753.499 n^6 + 49024.95 n^5
\nonumber \\
&+& 433177.7 n^4 + 1573332 n^3 + 961413.1 n^2 - 7398997 n -
1.495424~10^7 \biggr)
\nonumber \\
&+& {\tau^6 (n + 2) \over 64(n + 8)^{11}} \biggl(n^{10} + 102.140 n^9 +
4723.96 n^8 + 124384 n^7
\nonumber \\
&+& 1.93424~10^6 n^6 + 1.71104~10^7 n^5 + 8.17925~10^7 n^4 + 2.27733~10^8 n^3 +
\nonumber \\
&+& 5.19328~10^8 n^2 + 1.32120~10^9 n + 1.96458~10^9 \biggr),
\end{eqnarray}
\begin{eqnarray}
\label{alpha-tau}
\alpha &=& \frac{1}{2} - \frac{3 \tau (n + 2)}{4(n + 8)}
- {\tau^2 (n + 2) \over {72 (n + 8)^3}} \biggl(27 n^2 + 506 n + 1216
\biggr)
\nonumber \\
&-& {3 \tau^3 (n + 2)\over {16 (n + 8)^5}} \biggl(n^4 + 38.628325 n^3 +
456.42625 n^2 + 1330.8024 n + 460.64396 \biggr)
\nonumber \\
&-& {3 \tau^4 (n + 2)\over 32 (n + 8)^7} \biggl(n^6 + 59.190931 n^5 +
1377.3824 n^4 + 14095.696 n^3
\nonumber \\
&+& 56076.311 n^2 + 96550.551 n + 94371.957 \biggr)
\nonumber \\
&-& {3 \tau^5 (n + 2)\over 64 (n + 8)^9} \biggl(n^8 + 80.43442 n^7 +
2802.339 n^6 + 50883.40 n^5
\nonumber \\
&+& 468728.9 n^4 + 1911709 n^3 + 2598713 n^2 - 3400834 n -
1.087227~10^7 \biggr)
\nonumber \\
&-& {3 \tau^6 (n + 2)\over 128 (n + 8)^{11}} \biggl(n^{10} + 102.515 n^9
+ 4770.89 n^8 + 126952 n^7
\nonumber \\
&+& 2.01134~10^6 n^6 + 1.84139~10^7 n^5 + 9.40772~10^7 n^4 +  2.93115~10^8 n^3
\nonumber \\
&+& 7.17030~10^8 n^2 + 1.64131~10^9 n + 2.18006~10^9 \biggr). \ \
\end{eqnarray}

In the analysis of the experimental data, it is also important to know
the critical exponent of the scaling correction $\omega$, whose
pseudo-$\epsilon$ expansion has the form
\begin{eqnarray}
\label{omega-tau} \omega &=& \tau - {4\tau^2 \over 27(n + 8)^2} \biggl(41
n + 190 \biggr)
\nonumber \\
&+& {\tau^3 \over (n + 8)^4} \biggl(2.6978855 n^3 + 57.675086 n^2 +
594.43112 n + 1609.6102 \biggr)
\nonumber \\
&-& {\tau^4 \over (n + 8)^6} \biggl(-0.46693767 n^5 + 34.440983 n^4 +
1119.6990 n^3
\nonumber \\
&+& 11012.281 n^2 + 54280.495 n + 103646.63 \biggr)
\nonumber \\
&+& {\tau^5 \over (n + 8)^8} \biggl(0.2049447 n^7 + 20.18796 n^6 +
680.4944 n^5 + 17055.42 n^4
\nonumber \\
&+& 254491.6 n^3 + 1892609 n^2 + 6986639 n + 1.019704~10^7 \biggr)
\nonumber \\
&-& {\tau^6 \over (n + 8)^{10}} \biggl(-0.117121 n^9 - 12.1743 n^8 +
78.8694 n^7 + 21145.0 n^6
\nonumber \\
&+& 570921 n^5 + 7.69420~10^6 n^4 + 6.34720~10^7 n^3 + 3.19979~10^8 n^2
\nonumber \\
&+& 9.00151~10^8 n + 1.07362~10^9 \biggr). \ \
\end{eqnarray}
These $\tau$-series are used below to obtain the numerical values
of critical exponents for some special values of $n$.

\section{ Structure of the $\tau$-series and numerical results}

We consider pseudo-$\epsilon$ expansions (3)–-(5) for physically interesting
values of $n$. The most important cases are $n = 1$, $n = 2$ and $n = 3$,
which correspond to phase transitions in simple liquids and binary
mixtures, in easy-axis, easy-plane, and Heisenberg ferromagnets, in
superconductors with $s$-pairing, in Bose superfluids, and in many
other systems. The corresponding $\tau$-series have the forms
\begin{eqnarray}
\gamma &=& 1 + \frac{\tau}{6} + 0.061156836 \tau^{2} + 0.008519079
\tau^{3} + 0.00655499 \tau^{4}
\nonumber\\
&-& 0.00467841\tau^{5} + 0.0061748\tau^{6},
\end{eqnarray}
\begin{eqnarray}
\alpha &=& \frac{1}{2} - \frac{\tau}{4} - 0.099965706 \tau^2 -
0.02179070 \tau^3 - 0.01543751 \tau^4
\nonumber\\
&+& 0.0033540 \tau^5 - 0.011082 \tau^6,
\end{eqnarray}
\begin{eqnarray}
\omega &=& \tau - 0.42249657 \tau^2 + 0.34513249 \tau^3 - 0.32006015 \tau^4
\nonumber\\
&+& 0.4494775 \tau^5 - 0.678421 \tau^6
\end{eqnarray}
for $n = 1$,
\begin{eqnarray}
\gamma &=& 1+\frac{\tau}{5} + 0.080592593 \tau^2 + 0.01991018 \tau^3 +
0.01205280 \tau^4
\nonumber\\
&-& 0.00057920 \tau^5 + 0.0065646\tau^6,
\end{eqnarray}
\begin{eqnarray}
\alpha &=& \frac{1}{2} -\frac{3\tau}{10} - 0.129777778 \tau^2 -
    0.03954735 \tau^3 - 0.02432025 \tau^4
\nonumber\\
  &-& 0.0032498 \tau^5 - 0.012109 \tau^6,
\end{eqnarray}
\begin{eqnarray}
\omega &=& \tau - 0.40296296 \tau^2 + 0.30507558 \tau^3 - 0.26575046 \tau^4
\nonumber\\
&+& 0.3407266 \tau^5 - 0.480435 \tau^6
\end{eqnarray}
for $n = 2$, and
\begin{eqnarray}
\gamma &=& 1 + \frac{5}{22}\tau + 0.0974274425 \tau^2 + 0.03099116
\tau^3 + 0.01805659 \tau^4
\nonumber\\
&+& 0.00418601 \tau^5 + 0.0080091 \tau^6,
\end{eqnarray}
\begin{eqnarray}
\alpha &=& \frac{1}{2} -\frac{15\tau}{44} - 0.155323900 \tau^2 -
    0.05637685 \tau^3 - 0.03357917 \tau^4
\nonumber\\
  &-& 0.0105846 \tau^5 - 0.014517 \tau^6,
\end{eqnarray}
\begin{eqnarray}
\omega &=& \tau - 0.38322620 \tau^2 + 0.27216872 \tau^3 - 0.22494669 \tau^4
\nonumber\\
&+& 0.2641538 \tau^5 - 0.352583 \tau^6
\end{eqnarray}
for n = 3. It is physically interesting to consider the limit $n \to 0$,
where model (1) describes the critical behavior of polymers (self-avoiding walk).
In the case n = 0, the pseudo-$\epsilon$ expansions of the critical exponents become
\begin{eqnarray}
\gamma &=& 1 + \frac{\tau}{8} + 0.039351852 \tau^{2} - 0.00152232
\tau^{3} + 0.00269299 \tau^{4}
\nonumber\\
&-& 0.00696361 \tau^{5} + 0.0071471 \tau^{6},
\end{eqnarray}
\begin{eqnarray}
\alpha &=& \frac{1}{2} - \frac{3}{16} \tau - 0.065972222 \tau^2 -
0.00527165 \tau^3 - 0.00843751 \tau^4
\nonumber\\
&+& 0.0075942 \tau^5 - 0.011897 \tau^6,
\end{eqnarray}
\begin{eqnarray}
\omega &=& \tau - 0.43981481 \tau^2 + 0.39297124 \tau^3 - 0.39538053
\tau^4
\nonumber\\
&+& 0.6077908 \tau^5 - 0.999887 \tau^6.
\end{eqnarray}

We note that the pseudo-$\epsilon$ expansion method was previously
used to determine numerical values of critical exponents in \cite{LGZ80,GZJ98},
but no $\tau$-series (3)–-(17) were given in those papers. The resummation procedure
based on the Borel–Leroy transformation and the conformal map technique were
used there. Here, we determine the values of critical exponents without using
the Borel summation, which is a canonical tool in the theory of critical phenomena.

It follows from formulas (6)–-(17) that the $\tau$-series for the exponents
$\gamma$ and $\alpha$ have small coefficients rapidly decreasing in
absolute value. It is therefore natural to try to find the numerical
values of these exponents using the simplest way, namely, using Pad\'e
approximants and directly summing pseudo-$\epsilon$ expansions. It is
necessary to keep in mind that the obtained series, despite their favorable
structure, diverge in this case; the divergence, in particular, is manifested
by the behavior of the higher-order terms, which tend to increase. On the
other hand, because the initial RG expansions are asymptotic, there are reasons
to believe that the pseudo-$\epsilon$ expansions have the same property because
their coefficients can be obtained from the coefficients of the RG series by
finitely many algebraic operations with the coefficients of the same and lower
orders \cite{NS14e,NSHe}. This implies that in the case of direct summation of
some pseudo-$\epsilon$ expansion, the most precise estimate can be obtained by
calculating the partial sum of the series bounded by the term that is least
in absolute value.

The results of calculating the critical exponents of the susceptibility
and heat capacity using the Pad\'e approximants [L/M] and the efficiency
of this technique applied to pseudo-$\epsilon$ expansions are illustrated
in Tables 1–4, where the Pad\'e triangles are given for $\gamma$ and
$\alpha$ at $n = 1$ (the Ising model) and $n = 3$ (the Heisenberg model).
The bottom rows of these tables (RoC) show the character and the rate of
convergence of the Pad\'e estimates to the limit values. The $k$-th order
Pad\'e estimate is here the number given by the corresponding diagonal
approximant or by the half-sum of the values given by approximants of the
form [M/M-1] and [M-1/M] in the case without any diagonal approximant.

It can be seen that summing pseudo-$\epsilon$ expansions of the critical
exponents by the Pad\'e method generates an iteration procedure that
converges to the asymptotic values very rapidly. The following question
arises: How close are these values to the most precise numerical estimates
obtained by other methods? This question is answered in Tables 5 and 6,
where we present the values of the exponents $\gamma$ and $\alpha$ for
different $n$ obtained by the method described above ($\tau$, Pad\'e)
and their analogues obtained by resumming six-loop RG expansions by
the Pad\'e–Borel–Leroy (PBL) method and by the "conform-Borel" (CB) method,
by processing the strong coupling (SC) expansions, by processing the
five-loop $\epsilon$ expansions ($\epsilon$-exp), and by lattice
calculations (LC). These tables also contain the estimates of $\gamma$ and $\alpha$
obtained by optimally truncated direct summation of the pseudo-$\epsilon$ expansions,
i.e., of expansions truncated at the term that is least in absolute value
($\tau$, OTDS). To clarify the general properties of pseudo-$\epsilon$ expansions,
we calculated the critical exponents not only for physical values of $n$ but also
for $n > 3$. The corresponding results are also given in Tables 5 and 6.

It can be seen from these tables that the pseudo-$\epsilon$ expansion
method together with the Pad\'e approximant technique gives values of
critical exponents that are very close to their counterparts obtained
by alternative field theory methods and by lattice calculations. Moreover,
even the direct summation of pseudo-$\epsilon$ expansions of $\gamma$ and
$\alpha$ leads to numerical estimates that differ from the most reliable
values by their spread order, i.e., by the expected error of theoretical
determination of the exponents. The proximity between the numbers in the
first two rows in Tables 5 and 6 and the precise values obtained by different
methods confirms that the pseudo-$\epsilon$ expansion technique is numerically
highly efficient in such problems.

We further consider the critical exponent of the scaling correction
$\omega$. In this case, the computational situation is not so favorable as
in the case of the exponents $\gamma$ and $\alpha$. The coefficients of
pseudo-$\epsilon$ expansions (8), (11), (14), and (17) are not small and
do not exhibit any pronounced tendency to decrease. But the $\tau$-series
are sign-alternating and have a regular structure, i.e., their
coefficients are first monotonically decreasing (in absolute value) and
then monotonically increasing as their number grows. This allows
concluding that using Pad\'e approximants can also give reliable numerical
results in this case. Tables 7 and 8, where the Pad\'e triangles are given
for pseudo-$\epsilon$ expansions (8) and (14) and the summary Table 9
confirm this conclusion. We can see that the Pad\'e estimates of the
exponent $\omega$ agree well with the result of field theory and lattice
computations for all values of $n$ up to $n = 32$. At the same time, the
optimally truncated direct summation gives acceptable results in this case
only for sufficiently large $n$.

\section{Scaling and numerical efficiency}

Resumming the pseudo-$\epsilon$ expansions by the Pad\'e method, we can
calculate the critical exponents of three-dimensional systems with an
accuracy comparable to the accuracy of the most efficient lattice and
field theory iteration schemes. It is interesting to verify whether the
results thus obtained are intrinsically consistent. For this, we must use
expressions (3) and (4) to determine the pseudo-$\epsilon$ expansions of
the other critical exponents, sum the corresponding $\tau$-series in the
Pad\'e sense, and verify the extent to which the obtained numbers satisfy
the scaling relations. In this case, it is necessary to keep in mind that
the relation $\alpha = 2 - D\nu$ cannot be a source of the required
information, because it is exactly satisfied in our case for trivial
reasons. The formulas
\begin{eqnarray}
\nu = {\gamma \over {2 - \eta}}, \qquad \nu = {2\beta \over {1 + \eta}},
\qquad \alpha + 2\beta + \gamma = 2,
\end{eqnarray}
are efficient in this case, and we use them as test formulas. Table 10
shows the results of substituting the critical exponent values derived
from the corresponding $\tau$-series by Pad\'e resummations in these
formulas. We see that the pseudo-$\epsilon$ expansion technique reproduces
the scaling relations with an accuracy at the level of 0.005 and higher.
This confirms that the approach discussed here allows obtaining
high-precision numerical estimates even by the simplest resummation
methods.

What are the reasons for this very high numerical efficiency of the
pseudo-$\epsilon$ expansion method? We believe that the main
distinguishing characteristics of this method ensuring its computational
power are the following:

1. The coefficients of pseudo-$\epsilon$ expansions are significantly less
(in absolute value) and begin to increase much later than the coefficients
of the corresponding RG series. This can be explained by the structure of
the formulas relating the former to the latter. This structure ensures
multiple subtractions ("destructive interference") of the coefficients of
the initial RG expansions when the coefficients of the $\tau$-series are
calculated \cite{NS14e}.

2. The expressions for the coefficients of the kth-order pseudo-$\epsilon$
expansions contain not only the $k$-th order coefficients of the RG series
but also the RG coefficients of lower orders down to the first order
\cite{NSHe}. This means that the operations with pseudo-$\epsilon$
expansions use more information contained in the RG series than in the
case where the RG expansions themselves are processed.

3. In the pseudo-$\epsilon$ expansion method, the physical value of the
expansion parameter $\tau$ is equal to unity. The field theory RG
technique is based on the expansion in the Wilson fixed-point coordinate
$g$. We have $g \approx 1.4$ for three-dimensional systems with $0 \le n
\le 3$ and $g \approx 1.8$ for the two-dimensional Ising model, i. e., $g$
is significantly greater than unity. This difference is very significant,
especially in the case of high perturbation orders.

4. The pseudo-$\epsilon$ expansion technique also has some advantages over
the canonical Wilson–Fisher $\epsilon$ expansion method. The point is that
in the construction of the $\epsilon$ expansions, the integrals
corresponding to the Feynman diagrams are calculated in the $(4 -
\epsilon)$-dimensional space, while the coefficients of the
pseudo-$\epsilon$ expansions are expressed in terms of integrals
calculated for models with spaces of physical dimensions.

The results obtained above and the arguments listed above allow concluding
that the pseudo-$\epsilon$ expansion method can be regarded as a
distinctive resummation method that converts divergent RG series into
expansions very convenient for obtaining reliable numerical results.

This research is supported by St. Petersburg State University (Grant No.
11.38.636.2013), the Russian Foundation for Basic Research (Grant No.
15-02-04687), and the Dynasty Foundation.

\newpage

\begin{table}[t]
\caption{Pad\'e triangle for the pseudo-$\epsilon$ expansion of the
critical exponent $\gamma$ at $n = 1$. The approximant [3/1] has a pole
close to 1, which makes the corresponding value of $\gamma$ unreliable.
The bottom row (RoC) illustrates the character of convergence of the
Pad\'e estimates to the asymptotic value. Hereafter, the $k$-th order
Pad\'e estimate is the number given by the corresponding diagonal
approximant or the half-sum of the quantities given by approximants of the
form [M/M-1] and [M-1/M] in the case when diagonal approximant is absent.}
\label{tab1}
\renewcommand{\tabcolsep}{0.4cm}
\begin{tabular}{{c}|*{7}{c}}
$M \setminus L$ & 0 & 1 & 2 & 3 & 4 & 5 & 6 \\ \hline
0 & 1      & 1.1667 & 1.2278 & 1.2363 & 1.2429 & 1.2382 & 1.2444 \\
1 & 1.2000 & 1.2633 & 1.2377 & 1.2648 & 1.2402 & 1.2409 &        \\
2 & 1.2501 & 1.2408 & 1.2434 & 1.2412 & 1.2411 &        &        \\
3 & 1.2389 & 1.2430 & 1.2419 & 1.2411 &        &        &        \\
4 & 1.2455 & 1.2415 & 1.2409 &        &        &        &        \\
5 & 1.2358 & 1.2410 &        &        &        &        &        \\
6 & 1.2475 &        &        &        &        &        &        \\
\hline
RoC & 1    & 1.1833 & 1.2633 & 1.2393 & 1.2434 & 1.2416 & 1.2411 \\
\end{tabular}
\end{table}

\begin{table}[t]
\caption{Pad\'e triangle for the pseudo-$\epsilon$ expansion of the
critical exponent $\gamma$ at $n = 3$. The approximant [2/4] has a pole
close to 1, which makes the corresponding value of $\gamma$ unreliable.}
\label{tab2}
\renewcommand{\tabcolsep}{0.4cm}
\begin{tabular}{{c}|*{7}{c}}
$M \setminus L$ & 0 & 1 & 2 & 3 & 4 & 5 & 6 \\
\hline
0 & 1      & 1.2273 & 1.3247 & 1.3557 & 1.3737 & 1.3779 & 1.3859 \\
1 & 1.2941 & 1.3978 & 1.3701 & 1.3990 & 1.3792 & 1.3692 &        \\
2 & 1.3756 & 1.3728 & 1.3809 & 1.3829 & 1.3866 &        &        \\
3 & 1.3727 & 1.3751 & 1.3830 & 1.3796 &        &        &        \\
4 & 1.3858 & 1.3820 & 1.3999 &        &        &        &        \\
5 & 1.3805 & 1.3842 &        &        &        &        &        \\
6 & 1.3925 &        &        &        &        &        &        \\
\hline
RoC & 1    & 1.2607 & 1.3978 & 1.3715 & 1.3809 & 1.3829 & 1.3796 \\
\end{tabular}
\end{table}

\begin{table}[t]
\caption{Pad\'e triangle for the pseudo-$\epsilon$ expansion of the
critical exponent $\alpha$ at $n = 1$.} \label{tab3}
\renewcommand{\tabcolsep}{0.4cm}
\begin{tabular}{{c}|*{7}{c}}
$M \setminus L$ & 0 & 1 & 2 & 3 & 4 & 5 & 6 \\ \hline
0 & 0.5    & 0.25   & 0.1500 & 0.1282 & 0.1128 & 0.1162 & 0.1051 \\
1 & 0.3333 & 0.0834 & 0.1222 & 0.0753 & 0.1156 & 0.1136 & \\
2 & 0.2564 & 0.1254 & 0.1101 & 0.1119 & 0.1096 & & \\
3 & 0.2157 & 0.0959 & 0.1120 & 0.1109 & & & \\
4 & 0.1890 & 0.1232 & 0.1097 & & & &\\
5 & 0.1718 & 0.0777 & & & & & \\
6 & 0.1583 & & & & & &\\
\hline
RoC & 0.5  & 0.2917 & 0.0834 & 0.1238 & 0.1101 & 0.1120 & 0.1109 \\
\end{tabular}
\end{table}

\begin{table}[t]
\caption{Pad\'e triangle for the pseudo-$\epsilon$ expansion of the
critical exponent $\alpha$ at $n = 3$. The number in the bottom row (RoC)
marked by an asterisk is the estimate averaged over the approximants [4/2]
and [2/4] because the approximant [3/3] has a pole close to 1, which makes
the corresponding value of $\alpha$ unreliable. Hereafter, the results
obtained by averaging over the exponent values given by two off-diagonal
approximants are marked by asterisks.} \label{tab4}
\renewcommand{\tabcolsep}{0.3cm}
\begin{tabular}{{c}|*{7}{c}}
$M \setminus L$ & 0 & 1 & 2 & 3 & 4 & 5 & 6 \\
\hline
0 & 0.5    & 0.1591 & 0.0038 & $-0.0526$ & $-0.0862$ & $-0.0968$ & $-0.1113$ \\
1 & 0.2973 & $-0.1262$ & $-0.0847$ & $-0.1356$ & $-0.1016$ & $-0.0577$ & \\
2 & 0.2035 & $-0.0827$ & $-0.1029$ & $-0.1086$ & $-0.1157$ & & \\
3 & 0.1510 & $-0.1262$ & $-0.1084$ & $-0.0942$ & & & \\
4 & 0.1169 & $-0.0991$ & $-0.1145$ & & & &\\
5 & 0.0934 & $-0.1414$ & & & & & \\
6 & 0.0759 & & & & & &\\
\hline
RoC & 0.5  & 0.2282 & $-0.1262$ & $-0.0837$ & $-0.1029$ & $-0.1085$ & $-0.1151^*$ \\
\end{tabular}
\end{table}

\begin{table}[t]
\caption{Critical exponent $\gamma$ for different values of $n$ obtained
using Pad\'e approximants ($\tau$, Pad\'e) and the optimally truncated
direct summation of pseudo-$\epsilon$ expansions (3) ($\tau$, OTDS). The
numbers marked by asterisks are the estimates averaged over the
approximants [4/2] and [2/4] because the approximant [3/3] has poles close
to 1 in these cases. For comparison, we present the values of $\gamma$
obtained by resumming six-loop RG expansions using the Pad\'e-Borel-Leroy
(PBL) method and the "conform-Borel" (CB) method and the results obtained
by processing strong-coupling (SC) expansions, by processing five-loop
$\epsilon$ expansions ($\epsilon$-exp), and by lattice calculations (LC).
The "bc" note means that resumming an $\epsilon$ expansion is based on
using exact values of the exponents known for two-dimensional models. The
"sc" and "bcc" notes means that the results were obtained on simple cubic
and face-centered cubic lattices.} \label{tab5}
\begin{tabular}{*{9}{c}}
\hline
$n$~~&~~~0~~~&~~~1~~~&~~~2~~~&~~~3~~~&~~~4~~~&~~~8~~~&~~~16~~~&~~~32~\\
\hline $\tau$, Pade~       &  1.1617  &  1.2411  &  1.3154  & ~1.3796~ &
~$1.4572^*$~ & ~$1.6456^*$~ & ~1.8254~ & ~1.9138~ \\
$\tau$, OTDS~       & ~1.1628~ & ~1.2382~ & ~1.3120~ & ~1.3779~ & ~1.4358~ & ~1.6174~  & ~1.7769~ & ~1.8798~ \\
PBL \cite{BNM78}~ & ~1.161~  & ~1.241~  & ~1.316~  & ~1.39~   &          &           &          &          \\
CB \cite{LGZ80}~  & ~1.1615~ & ~1.241~  & ~1.316~  & ~1.386~  &          &           &          &          \\
PBL \cite{AS95}~  & ~1.160~  & ~1.239~  & ~1.315~  & ~1.386~  & ~1.449~  & ~1.637~   & 1.807    & 1.908    \\
CB \cite{GZJ98}~  & ~1.1596~ & ~1.2396~ & ~1.3169~ & ~1.3895~ & ~1.456~  &           &          &          \\
SC \cite{Kl99}~   & ~1.161~  & ~1.241~  & ~1.318~  & ~1.390~  & ~1.451~  & ~1.638~   & 1.822    & 1.920    \\
CB \cite{Susl08}~ & ~1.1615~ & ~1.2411~ & ~1.3172~ & ~1.3876~ &          &           &          &          \\
$\epsilon$-exp \cite{GZJ98}~ & ~1.1575~ & ~1.2355~ & ~1.311~  & ~1.382~ & ~1.448~ &   &          &          \\
$\epsilon$-exp \cite{GZJ98}~ & ~1.1571~ & ~1.2380~ & ~1.317~  & ~1.392~ & ~1.460  &       &      &          \\
LC \cite{BC97}      & ~1.1594~ & ~1.2388~ & ~1.325~  & ~1.406~  &  &  &  &   \\
LC \cite{BC97}      & ~1.1582~ & ~1.2384~ & ~1.322~  & ~1.402~  &  &  &  &   \\
LC                  & ~1.1575 \cite{Car98}~ & ~1.2372 \cite{H10}~ &
~1.3177 \cite{CHPRV2001}~ & ~1.3960 \cite{CHPRV2002}~  &  & 1.68 \cite{DPV15}  &   &   \\
LC                  & ~1.1573 \cite{PV07}~ & ~1.2373 \cite{CPRV2002}~ & ~1.3178 \cite{CHPV2006}~ &  &  &  &  &  \\
\hline
\end{tabular}
\end{table}

\begin{table}[t]
\caption{Critical exponent $\alpha$ for different values of $n$ obtained
using Pad\'e approximants ($\tau$, Pad\'e) and the optimally truncated
direct summation of pseudo-$\epsilon$ expansions (4) ($\tau$, OTDS). The
numbers marked by asterisks were obtained by averaging the values given by
the approximants [4/2] and [2/4]; the diagonal approximant [3/3] has poles
close to 1 in these cases. For comparison, we present the values of
$\alpha$ obtained by resumming six-loop RG expansions by the
Pad\'e-Borel-Leroy (PBL) method and the "conform-Borel" (CB) technique and
the results obtained by processing five-loop $\epsilon$ expansions
($\epsilon$-exp) and by lattice calculations (LC). The "bc," "sc," and
"bcc" notes are the same as in Table 5.} \label{tab6}
\begin{tabular}{*{9}{c}}
\hline
$n$~~&~~~0~~~&~~~1~~~&~~~2~~~&~~~3~~~&~~~4~~~&~~~8~~~&~~~16~~~&~~~32~\\
\hline
$\tau$, Pade~       &  0.2355  &  0.1109  & $-0.0023$  &
$-0.1151^*$ & $-0.2105^*$ & $-0.4828^*$ & $-0.7181^*$ & $-0.8841$ \\
$\tau$, OTDS~       & ~0.2413~ &  0.1162  &  0.0031 & $-0.0968$ & $-0.2011$ & $-0.4549$ & $-0.6859$ & $-0.8319$ \\
PBL \cite{BNM78}~ & ~0.236~  & ~0.110~  & $-0.007$ & $-0.115$ &          &           &           &          \\
CB \cite{LGZ80}~  & ~0.236~  & ~0.110~  & $-0.007$ & $-0.115$ &          &           &           &          \\
PBL \cite{AS95}~  & ~0.231~  & ~0.107~  & $-0.010$ & $-0.117$ & $-0.213$ & $-0.489$ & $-0.732$  & $-0.875$ \\
CB \cite{GZJ98}~  & ~0.235~  & ~0.109~  & $-0.011$ & $-0.122$ & $-0.223$ &           &           &          \\
$\epsilon$-exp \cite{GZJ98}~  & ~0.2375~ & ~0.1130~& $-0.0040$ & $-0.1135$ & $-0.211$ &       &          &  \\
$\epsilon$-exp \cite{GZJ98}, bc & ~0.2366~ & ~0.1085~& $-0.013$ & $-0.124$ & $-0.226$ &       &          &  \\
LC \cite{BC99}, sc & ~0.24~ & ~0.103~ & $-0.014$~ & $-0.11$~ & $-0.22$ &   &   &   \\
LC \cite{BC99}, bcc  & ~0.235~ & ~0.105~ & $-0.019$~ & $-0.13$~ & $-0.25$ &   &   &   \\
LC  & ~0.2370 & ~0.1096 & $-0.0151$ & $-0.1336$   \\
  & \cite{PV07}~ & \cite{CPRV2002}~ & \cite{CHPV2006}~& \cite{CHPRV2002}
& & & & \\
\hline
\end{tabular}
\end{table}

\begin{table}[t]
\caption{Pad\'e triangle for the pseudo-$\epsilon$ expansion of the
critical exponent $\omega$ for $n = 1$. The Pad\'e approximants [L/M] are
calculated for the ratio $\omega/\tau$, i.e., in the case where the
trivial multiplier $\tau = 1$ is neglected.} \label{tab7}
\renewcommand{\tabcolsep}{0.4cm}
\begin{tabular}{{c}|*{6}{c}}
$M \setminus L$ & 0 & 1 & 2 & 3 & 4 & 5  \\ \hline
0 & 1      & 0.5775 & 0.9226 & 0.6026 & 1.0521 & 0.3736 \\
1 & 0.7030 & 0.7675 & 0.7566 & 0.7895 & 0.7817 &        \\
2 & 0.7963 & 0.7577 & 0.7652 & 0.7830 &        &        \\
3 & 0.7355 & 0.7752 & 0.7906 &        &        &        \\
4 & 0.8720 & 0.7856 &        &        &        &        \\
5 & 0.6867 &        &        &        &        &        \\
\hline
RoC & 1    & 0.6402 & 0.7675 & 0.7571 & 0.7652 & 0.7868 \\
\end{tabular}
\end{table}

\begin{table}[t]
\caption{Pad\'e triangle for the pseudo-$\epsilon$ expansion of the
critical exponent $\omega$ for $n = 3$. The Pad\'e approximants [L/M] are
obtained for the ratio $\omega/\tau$, i.e., in the case where the
multiplier $\tau$ is neglected.} \label{tab8}
\renewcommand{\tabcolsep}{0.4cm}
\begin{tabular}{{c}|*{6}{c}}
$M \setminus L$ & 0 & 1 & 2 & 3 & 4 & 5  \\
\hline
0 & 1      & 0.6168 & 0.8889 & 0.6640 & 0.9281 & 0.5756 \\
1 & 0.7229 & 0.7759 & 0.7658 & 0.7855 & 0.7771 &        \\
2 & 0.7950 & 0.7669 & 0.7729 & 0.7793 &        &        \\
3 & 0.7516 & 0.7777 & 0.7807 &        &        &        \\
4 & 0.8234 & 0.7803 &        &        &        &        \\
5 & 0.7281 &        &        &        &        &        \\
\hline
RoC & 1    & 0.6699 & 0.7759 & 0.7664 & 0.7729 & 0.7800 \\
\end{tabular}
\end{table}

\begin{table}[t]
\caption{Critical exponent $\omega$ for different values of $n$ obtained
by resumming expansion (5) by using Pad\'e approximants ($\tau$, Pad\'e)
and by optimally truncated direct summation ($\tau$, OTDS). For
comparison, we present the values of $\omega$ obtained by processing
six-loop RG expansions by the Pad\'e-Borel-Leroy (PBL) method and the
"conform-Borel" (CB) technique and the results obtained by processing
strong coupling (SC) expansions, by processing five-loop $\epsilon$
expansions ($\epsilon$-exp), and by lattice calculations (LC).}
\label{tab9}
\begin{tabular}{*{9}{c}}
\hline
$n$~~&~~~0~~~&~~~1~~~&~~~2~~~&~~~3~~~&~~~4~~~&~~~8~~~&~~~16~~~&~~~32~\\
\hline
$\tau$, Pade~ & ~~0.7947~~ & ~~0.7868~~ & ~~0.7812~~
&~~0.7800~~ & ~~0.7825~~ & ~~0.8082~~ & ~~0.8607~~ & ~~0.9201~~ \\
$\tau$, OTDS~       & 0.9532 & 0.6026 & 0.6364 & 0.6640 & 0.6877 & 0.8557 & 0.8478 & 0.9181 \\
PBL \cite{BNM78}~ & 0.794 & 0.788 & 0.78  & 0.78  &       &       &       &        \\
CB \cite{LGZ80}~  & 0.80  & 0.79  & 0.78  & 0.78  &       &       &       &        \\
PBL \cite{S98}~   &       & 0.781 & 0.780 & 0.780 & 0.783 & 0.808 & 0.861 & 0.919  \\
CB \cite{GZJ98}~  & 0.812 & 0.799 & 0.789 & 0.782 & 0.774 &       &       &        \\
SC \cite{Kl99}~   & 0.810 & 0.805 & 0.800 & 0.797 & 0.795 & 0.810 & 0.862 & 0.924  \\
CB \cite{Susl08}~ & 0.790 & 0.782 & 0.778 & 0.778 &       &       &       &        \\
$\epsilon$-exp \cite{GZJ98} & 0.828  & 0.814  & 0.802  & 0.794 & 0.795 &       &     &  \\
$\epsilon$-exp \cite{LGZ85} & 0.82   & 0.81   & 0.80   & 0.79  &       &        &     &  \\
LC                   &  & ~0.83 \cite{CPRV2002}~ &~0.785 \cite{CHPV2006}~
& 0.773 \cite{H2001} &
0.765 \cite{H2001} &  &  &  \\
\hline
\end{tabular}
\end{table}

\begin{table}[t]
\caption{Accuracy of scaling relations (18) for the critical exponents
obtained by resumming the corresponding pseudo-$\epsilon$ expansions using
Pad\'e approximants.}
 \label{tab10}
\renewcommand{\tabcolsep}{0.15cm}
\begin{tabular}{{c}|*{8}{c}}
\hline
$n$ & 0 & 1 & 2 & 3 & 4 & 8 & 16 & 32 \\
\hline
${{\gamma \over{2-\eta}}-\nu}$ & 0.0009 & 0.0012 & 0.0017 & 0.0037   & $0.0014^*$ & $0.0010^*$ & $-0.0006$ & $-0.0001$ \\
                             &        &        &        & $0.0019^*$ & 0.0142   &          &           &  \\
\hline
${2\beta \over {1+\eta}}-\nu$ & $-0.0021$ & $-0.0024$ & $-0.0033$  & $-0.0055$   & $-0.0009^*$ & $-0.0015^*$ & 0.0040 & 0.0007 \\
                              &           &           &             & $-0.0006^*$ & $-0.0076$  &             &        &   \\
\hline
$\beta+{{\alpha+\gamma}\over 2}-1$ & $-0.0002$ & $-0.0001$ & 0.0000 & 0.0008   & $0.0047^*$  & $0.0079^*$ & 0.0014 & 0.0003 \\
                                   &           &           &        & $0.0042^*$ & $-0.0101$ &          &        &  \\
\hline
\end{tabular}
\end{table}


\begin{thebibliography}{999}

\bibitem{BNGM76} G. A. Baker, B. G. Nickel, M. S. Green, D. I. Meiron, ~Phys. Rev.
Lett. {\bf 36}, 1351 (1976).

\bibitem{Guelph} B. G. Nickel, D. I. Meiron, G. A. Baker, University of Guelph Report, 1977.

\bibitem{BNM78} G. A. Baker, B. G. Nickel, D. I. Meiron, ~Phys. Rev. B {\bf 17}, 1365 (1978).

\bibitem{MN91} D. B. Murray, B. G. Nickel, University of Guelph preprint, 1991.

\bibitem{AS95} S. A. Antonenko, A. I. Sokolov, ~Phys. Rev. E {\bf 51} 1894
(1995), arXiv: hep-th/9803264.

\bibitem{LGZ80} J. C. Le Guillou, J. Zinn-Justin, ~Phys. Rev. B {\bf 21}, 3976 (1980).

\bibitem{GZJ98} R. Guida, J. Zinn-Justin, ~J. Phys. A {\bf 31}, 3976
(1998), arXiv: cond-mat/9803240.

\bibitem{Kl99} H. Kleinert, Phys. Rev. D {\bf 60}, 850001 (1999), arXiv: hep-th/9812197.

\bibitem{Susl08} A. A. Pogorelov, I. M. Suslov, Zh. Eksp. Teor. Fiz. {\bf 133}, 1277
(2008) [JETP {\bf 106}, 1118 (2008)], arXiv:1010.3389.

\bibitem{ZJ01}  J. Zinn-Justin, ~Phys. Reports {\bf 344}, 159 (2001), arXiv: hep-th/0002136.

\bibitem{ZJ} J. Zinn-Justin. Quantum Field Theory and Critical Phenomena.
Clarendon Press, Oxford (2002).

\bibitem{KS01} H. Kleinert, V. Schulte-Frohlinde. Critical Properties of $\phi^4$
Theories. World Scientific, Singapore (2001).

\bibitem{PV02}  A. Pelissetto, E. Vicari, ~Phys. Reports {\bf 368}, 549
(2002), arXiv: cond-mat/0012164.

\bibitem{Susl05} I. M. Suslov, Zh. Eksp. Teor. Fiz. {\bf 127}, 1350 (2005) [JETP {\bf 100},
1188 (2005)], arXiv: hep-ph/0510142.

\bibitem{FH97} C. von Ferber, Yu. Holovatch, Phys. Rev. E {\bf 56}, 6370
(1997), arXiv: cond-mat/9705278.

\bibitem{FH99} C. von Ferber, Yu. Holovatch, Phys. Rev. E {\bf 59}, 6914
(1999), arXiv: cond-mat/9812119.

\bibitem{FHY2000} R. Folk, Yu. Holovatch, T. Yavorskii, Phys. Rev. B {\bf 62}, 12195 (2000),
arXiv: cond-mat/0003216.

\bibitem{HID04} Yu. Holovatch, D. Ivaneiko, B. Delamotte, J. Phys. A {\bf 37}, 3569 (2004),
arXiv: cond-mat/0312260.

\bibitem{CP05} P. Calabrese, P. Parruccini, Phys. Rev. B {\bf 71}, 064416 (2005), arXiv: cond-mat/0411027.

\bibitem{NS14c} A. I. Sokolov, M. A. Nikitina, Phys. Rev. E {\bf 89}, 052127 (2014), arXiv:1402.3531.

\bibitem{NS14e} A. I. Sokolov, M. A. Nikitina, Phys. Rev. E {\bf 90}, 012102 (2014), arXiv:1402.3894.

\bibitem{COPS04} P. Calabrese, E. V. Orlov, D. V. Pakhnin, A. I. Sokolov, Phys. Rev. B {\bf 70},
094425 (2004), arXiv: cond-mat/0405432.

\bibitem{S05} A. I. Sokolov, Fiz. Tverd. Tela {\bf 47}, 2056 (2005) [(Phys. Sol. State {\bf 47},
2144 (2005)], arXiv: cond-mat/0510088.

\bibitem{S13} A. I. Sokolov, Teor. Mat. Fiz. {\bf 176}, 140 (2013)[(Theor. Math. Phys. {\bf 176},
948 (2013)].

\bibitem{NS13} M. A. Nikitina, A. I. Sokolov, Phys. Rev. E {\bf 89}, 042146 (2014), arXiv:1312.1062.

\bibitem{NSHe} A. I. Sokolov, M. A. Nikitina, Physica A {\bf 444}, 177 (2016), arXiv:1402.4318.

\bibitem{BC97}  P. Butera and M. Comi, ~Phys. Rev. B {\bf 56}, 8212 (1997), arXiv: hep-lat/9703018.

\bibitem{Car98} S. Caracciolo, M. S. Causo, E. Vicari, Phys. Rev. E {\bf 57}, R1215
(1998), arXiv: cond-mat/9703250.

\bibitem{H10} M. Hasenbusch, Phys. Rev. B {\bf 82}, 174433 (2010), arXiv:1004.4486.

\bibitem{CHPRV2001} M. Campostrini, M. Hasenbusch, A. Pelissetto, P. Rossi, E. Vicari,
Phys. Rev. B {\bf 63}, 214503 (2001), arXiv: cond-mat/0010360.

\bibitem{CHPRV2002} M. Campostrini, M. Hasenbusch, A. Pelissetto, P. Rossi, E. Vicari,
Phys. Rev. B {\bf 65}, 144520 (2002), arXiv: cond-mat/0110336.

\bibitem{DPV15}  F. Delfino, A. Pelissetto, E. Vicari, Phys. Rev. E {\bf 91}, 052109 (2015),
arXiv:1502.07599.

\bibitem{PV07}  A. Pelissetto, E. Vicari, ~J. Phys. A {\bf 40}, F539 (2007),
arXiv: cond-mat/0703114.

\bibitem{CPRV2002} M. Campostrini, A. Pelissetto, P. Rossi, E. Vicari,
Phys. Rev. E {\bf 65}, 066127 (2002), arXiv: cond-mat/0201180.

\bibitem{CHPV2006} M. Campostrini, M. Hasenbusch, A. Pelissetto, E. Vicari,
Phys. Rev. B {\bf 74}, 144506 (2006), arXiv: cond-mat/0605083.

\bibitem{BC99}  P. Butera and M. Comi, ~Phys. Rev. B {\bf 60}, 6749 (1999),
arXiv: hep-lat/9903010.

\bibitem{S98}  A. I. Sokolov, Fiz. Tverd. Tela {\bf 40}, 1284 (1998)
Phys. Solid State {\bf 40}, 1169 (1998).

\bibitem{LGZ85} J. C. Le Guillou, J. Zinn-Justin, J. Physique Lett. {\bf 46}, 137 (1985).

\bibitem{H2001}  M. Hasenbusch, ~J. Phys. A {\bf 34}, 8221 (2001), arXiv: cond-mat/0010463.

\end{thebibliography}
\end{document}